\documentclass[final,3p,11pt,twocolumn]{elsarticle}

\usepackage{epsfig}
\usepackage{amssymb}
\usepackage{amsmath}
\usepackage{relsize}

\expandafter\let\csname equation*\endcsname\relax
\expandafter\let\csname endequation*\endcsname\relax 
\usepackage{amsmath}
\usepackage{amssymb}

\biboptions{compress}
\journal{Physics Letters A}

\def\vecb#1{\boldsymbol{#1}}

\def\abs#1{\left\lvert#1\right\rvert}

\newcommand{\ud}{d}

\def\abs#1{\left|#1\right|}

\begin{document}

\begin{frontmatter}

\title{Classification of excited-state quantum phase transitions for arbitrary number of degrees of freedom}

\author{Pavel Str{\'a}nsk{\'y}}\corref{cor} 
\author{Pavel Cejnar}
\address{Institute of Particle and Nuclear Physics, Faculty of Mathematics and Physics, Charles University, 
  V~Hole{\v s}ovi{\v c}k{\' a}ch 2, 180\,00 Prague, Czech Republic}

\cortext[cor]{The corresponding author; email address:\\ stransky@ipnp.troja.mff.cuni.cz}

\begin{abstract}
Classical stationary points of an analytic Hamiltonian induce singularities of the density of quantum energy levels and their flow with a control parameter in the system's infinite-size limit.
We show that for a system with $f$ degrees of freedom, a non-degenerate stationary point with index $r$ causes a discontinuity (for $r$ even) or divergence ($r$ odd) of the $(f$$-$$1)$\,th derivative of both density and flow of the spectrum. 
An increase of flatness for a degenerate stationary point shifts the singularity to lower derivatives.
The findings are verified in an $f=3$ toy model.
\end{abstract}



\end{frontmatter}

\section{Introduction}
\label{sec:Int}

Excited-State Quantum Phase Transitions (ESQPTs) are singularities observed in discrete energy spectra of some bound quantum systems in the infinite-size limit \cite{Cej06,Cap08,Cej08}.
They show up as non-analyticities in the density of quantum energy eigenstates as a function of energy $E$ and in the flow of the excited spectrum with a suitable control parameter $\lambda$.
The ESQPT critical borderlines in the $\lambda\times E$ plane are usually terminated by critical points of the ground-state Quantum Phase Transitions (QPTs) \cite{Sac99,Car10}, so they can be seen as extensions of the QPTs to the excited domain.
Thermodynamical and dynamical consequences of ESQPTs, as well as their experimental evidence in some synthetic quantum systems are currently focus of intense research, see {\it e.g.} Refs.\,\cite{Die13, Bra13,Bas14,San15}.

The ESQPT singularities in systems with a single effective degree of freedom, $f=1$, are most dramatic and have been known for long, see {\it e.g.} Refs.\,\cite{Ley05,Rib08,Rel08,Fer11}.
For increasing numbers of degrees of freedom $f$, the ESQPTs affect higher and higher derivatives of the level density and flow of the spectrum.
Their effects in systems with $f=2$ have been thoroughly studied in our recent works \cite{Str14,Str15}.
These analyses contain the prerequisites for an ESQPT theory in an arbitrary number of degrees of freedom, but only for systems whose Hamiltonian is of the form
\begin{equation}
\label{eq:Hstand}
H=\frac{\vecb{p}^{2}}{2}+V(\vecb{q})
\,,
\end{equation}
where $V(\vecb{q})$ is an analytic potential depending on coordinates $\vecb{q}$ and $\vecb{p}^{2}/2$ is a coordinate-independent kinetic energy, which is quadratic in momenta $\vecb{p}$. 
In this case, ESQPTs appear at energies corresponding to stationary points of $V(\vecb{q})$ above the main minimum, the corresponding defects in the spectrum being related to the stationary-point types.

The aim of the present paper is to develop a general-$f$ ESQPT theory for systems with unrestricted forms of the Hamilton function $H(\vecb{q},\vecb{p})$.
It should be stressed that Hamiltonians with non-trivial couplings between coordinates and momenta are common in algebraic models of many-body collective dynamics because generators of the corresponding dynamical groups are usually formed by combinations of coordinate and momentum operators \cite{Zha90}.
We develop a full classification of ESQPTs caused by non-degenerate (quadratic) stationary points of a general Hamiltonian.
Although our approach is rooted in the evaluation of the system's level density, we show that non-analyticities in this quantity affect correspondingly the flow properties of the spectrum with a variable parameter (the \lq\lq level dynamics\rq\rq).
Our conclusions are illustrated by a simple model.
An example of an ESQPT due to a degenerate stationary point is also analyzed, although for such higher-order stationary points no general classification exists.

The paper is organized as follows:
Section~\ref{sec:Den} analyzes the semiclassical level density in a vicinity of a non-degenerate stationary point and exemplifies a degenerate case.
Section~\ref{sec:Flora} aims at the impact of stationary points on flow properties of the spectrum. 
Section~\ref{sec:Model} presents a toy model with $f=3$.
Section~\ref{sec:Conclusion} brings a brief summary.

\section{Level density}
\label{sec:Den}

The quantum level density for a system with discrete energy spectrum is defined by
\begin{equation}
\rho(E)=\sum_{l}\delta(E-E_l)
\label{de}
\,,
\end{equation}
where $\delta$ stands for the Dirac function and $E_0\leq E_1\leq E_2,\dots$ denote individual energy eigenvalues.
The level density can be decomposed into a sum of smooth and oscillatory components:
\begin{equation}
\label{eq:rho}
\rho(E)=\bar{\rho}(E)+\widetilde{\rho}(E)
\,.
\end{equation}
The smooth component $\bar{\rho}$ captures the mean energy dependence of the level density (obtained, {\it e.g}, by a convolution of $\rho$ with a sufficiently wide smoothening function), while the oscillatory component $\widetilde{\rho}$ has a zero mean and collects fluctuations (the balance between the smoothed dependence and the full discrete spectrum).

In the limit $\hbar\to 0$, which in systems with finite numbers of degrees of freedom $f$ is equivalent to the infinite size limit \cite{Str14,Yaf82}, the fluctuations of the level density become infinitely dense and the oscillatory component gets washed out even by smoothening over an infinitesimal energy interval.
We shall therefore focus on the smooth component only. 
It can be determined from the size of the accessible phase space at a given energy:
\begin{equation}
\label{eq:rhobar}
\bar{\rho}(E)=\left(\frac{1}{2\pi\hbar}\right)^{f}
\underbrace{\int\ud^{2f}\vecb{x}\,\delta(E-H(\vecb{x}))}
_{\frac{\partial}{\partial E}\underbrace{\smallint_{H(\vecb{x})\leq E} d^{2f}\vecb{x}}_{\Omega(E)}}
\,,
\end{equation}
where $\vecb{x}\equiv(\vecb{p},\vecb{q})$ is a $2f$-dimensional vector containing $f$-dimensional vectors of coordinates $\vecb{q}$ and momenta $\vecb{p}$, and $H(\vecb{x})$ is the classical Hamiltonian of the system.
Note that Eq.\,\eqref{eq:rhobar} can be derived by the Feynman integration over the orbits of zero length, while the oscillatory component is analogously linked to classical periodic orbits \cite{Sto99}. 

\subsection{Effects of stationary points}
\label{sec:smoo}

As seen in Eq.\,\eqref{eq:rhobar}, the smooth component of the level density is proportional to the energy derivative of the volume function $\Omega(E)$ associated with the Hamiltonian $H(\vecb{x})$.
The function $\Omega(E)$ measures the $2f$-dimensional volume of the phase--space region satisfying $H(\vecb{x})\leq E$.
Even for analytic classical Hamiltonian forms $H(\vecb{x})$ it develops singularities at the points where the $(2f$$-$$1)$-dimensional hypersurface determined by the constant energy condition $H(\vecb{x})=E$ crosses a stationary point of $H$.
Indeed, using the substitution formula $\delta(\chi(\vecb{x}))=\sum_i\delta(\vecb{x}-\vecb{x}_{0i})/|\vecb{\nabla}_n\chi(\vecb{x}_{0i})|$,
where $\chi$ is any function satisfying $\chi(\vecb{x}_{0i})=0$ $\forall i$ and $\vecb{\nabla}_n$
stands for $n$-dimensional gradient, we express Eq.\,\eqref{eq:rhobar} via integration of the reciprocal gradient $1/|\vecb{\nabla}_{2f}H|$ over the constant-energy hypersurface. 
Therefore, the smooth component of the level density has non-analyticities at energies $E_{\vecb{w}}\equiv H(\vecb{w})$ corresponding to points $\vecb{w}$ where $\vecb{\nabla}_{2f}H(\vecb{w})=0$.

The impact of a stationary point of a given type on the level density has a universal character---it depends only on the local behavior of $H(\vecb{x})$ near $\vecb{w}$, and not on the global, system-specific features of dynamics.
This conclusion can be verified by a splitting of the integral in Eq.\,\eqref{eq:rhobar} near the stationary-point energy $E_{\vecb{w}}$ into a sum of regular and irregular parts:
\begin{equation}
\label{eq:rhobar2}
\bar{\rho}(E)=\bar{\rho}_{\vecb{w}}^{(0)}(E)+\bar{\rho}_{\vecb{w}}(E)
\,.
\end{equation}
The irregular part $\bar{\rho}_{\vecb{w}}$ contains integration over a small phase-space neighborhood of the stationary point and captures all non-analytic behavior of $\bar{\rho}$ due to the stationary point.
The regular part $\bar{\rho}_{\vecb{w}}^{(0)}$ contains the integration over the rest of the accessible phase space and yields an analytic contribution to $\bar{\rho}$.
To classify the singularity in the level density caused by the stationary point, it is sufficient to analyze properties of the irregular part.

Singularities of various volume functions have been studied in rather different contexts.
For instance, the so-called level set method of computational geometry and image processing (see e.g. Ref.\,\cite{LSM}) describes the motion of a general interface $\Gamma$ in an $n$-dimensional space of variable $\vecb{x}$ via a suitably determined function $\varphi(\vecb{x},t)$ such that $\Gamma(t)$ at any time $t$ coincides with the set of $\vecb{x}$ with $\varphi=0$.
In this method, an expression analogous to Eq.\,\eqref{eq:rhobar} represents time derivative of the volume $\Omega$ bounded by $\Gamma$.
Non-differentiability of the volume function at the places where the interface crosses a non-degenerate stationary point of $\varphi$ has been analyzed by Hoveijn \cite{Hov08}.
A similar problem has been addressed also by Kastner {\it et al.} \cite{Kas08a,Kas08b,Kas08c} in the framework of statistical mechanics, namely in connection with the so-called configurational state density defined by $\omega(E)=\int\ud^{f}\vecb{q}\,\delta(E-V(\vecb{q}))$, in analogy to Eq.\,\eqref{eq:rhobar}, for systems with Hamiltonians of the type \eqref{eq:Hstand}.
Mathematical conditions have been formulated for the occurrence of a thermodynamic phase transition caused by stationary points of the potential energy landscape \cite{Kas08b}.

Our aim in this paper is to analyze the link of various classical stationary points to the ESQPT singularities in quantal spectra of general Hamiltonian systems.
At first we show (Sec.\,\ref{sec:ndg}) that there exists a finite classification of ESQPTs corresponding to non-degenerate (quadratic) stationary points of various types, where \lq\lq classification\rq\rq\ means typology of discontinuities or singularities that for a given number of degrees of freedom $f$ appear in the $(f$$-$$1)$\,th energy derivative of $\bar{\rho}$. 
The results in this part are mathematically equivalent to those derived previously in other contexts \cite{Hov08,Kas08a,Kas08b,Kas08c}.
Second, although a classification of ESQPTs caused by higher-order stationary points does not exist, we give an illustrative example of this kind (Sec.\,\ref{sec:dg}) and discuss conditions for descending the ESQPT signatures to lower derivatives of $\bar{\rho}$.
Third, we analyze (Sec.\,\ref{sec:Flora}) generic effects of stationary points on the flow of quantum spectrum with the Hamiltonian control parameters, showing that a smoothed flow is generally expected to exhibit the same type of singularity as level density.

\subsection{Singularities caused by non-degenerate stationary points}
\label{sec:ndg}

The analysis presented in this section is based on the Morse theory~\cite{Mat02}.
Let ${\cal M}$ be an $n$-dimensional manifold of points $\vecb{x}=(x_1,\dots,x_n)$ and $H(\vecb{x})$ a smooth function $H:{\cal M}\rightarrow\mathbb{R}$.
Consider a stationary point $\vecb{w}$ satisfying $\vecb{\nabla}_{n}H(\vecb{w})=0$.
The stationary point is called non-degenerate if the Hessian matrix $\mathcal{H}(\vecb{w})$ with elements $\mathcal{H}_{ij}(\vecb{w})=\partial^{2}H(\vecb{x})/(\partial x_{i}\partial x_{j})|_{\vecb{x}=\vecb{w}}$ has only non-zero eigenvalues.
This means that the function $H(\vecb{x})$ is locally quadratic in all directions at the point $\vecb{w}$.
According to the Morse lemma~\cite{Mat02}, close to any non-degenerate stationary point $\vecb{w}$ one can choose coordinates $\vecb{y}$ such that the following equality holds up to the ${\cal O}(y_i^3)$ terms:
\begin{align}
\label{eq:HMorse}
H_{\vecb{w}}(\vecb{y})
&=H(\vecb{w})
\\
&
\underbrace{-y_{1}^{2}-\dotsb-y_{r}^{2}}_{-R_-^2}
\underbrace{+y_{r+1}^{2}+\dotsb+y_{n}^{2}}_{+R_+^2}
\nonumber
\,.
\end{align}
The integer $r$, called the \emph{index} of stationary point, is equal to the number of negative eigenvalues of 
$\mathcal{H}(\vecb{w})$.
Variable $R_{-}$ is a radial coordinate in the $r$-dimensional subspace connected with negative eigenvalues of the Hessian matrix, while $R_{+}$ is a radial coordinate in the adjunct $s$-dimensional subspace ($s=2f-r$) connected with the positive eigenvalues.

The Morse theory can be applied to system with $f$ degrees of freedom.
Let $H(\vecb{x})\equiv H(\vecb{p},\vecb{q})$ be Hamilton function in the phase space of dimension $n=2f$.
We stress that $H$ is allowed to mix coordinates $\vecb{q}$ and momenta $\vecb{p}$ in an arbitrary way.
In a vicinity of a non-degenerate stationary point $\vecb{w}$, the Hamiltonian can be locally expressed in the form (\ref{eq:HMorse}).
At the energy of the stationary point, $E=E_{\vecb{w}}\equiv H(\vecb{w})$, the smooth component of the level density develops a non-analyticity.
The irregular part $\bar{\rho}_{\vecb{w}}$ of $\bar{\rho}$ in a vicinity of $E_{\vecb{w}}$ can be evaluated explicitly.
In particular, the integration in Eq.\,\eqref{eq:rhobar} close to the stationary point can be performed with the aid of the expansion (\ref{eq:HMorse}), separately in the $r$- and $s$-dimensional subspaces of negative and positive Hessian eigenvalues, respectively.

This method yields the irregular part of the level density at energy $E=E_{\vecb{w}}+\Delta$, where $|\Delta|$ is sufficiently small, in the form of the following expressions:
\begin{equation}
\label{eq:rho0}
\bar{\rho}_{\vecb{w}}(\Delta)=
\left(\tfrac{1}{2\pi\hbar}\right)^{f}
\!\!\!
\tfrac{\sigma_s}{2\det^{1/2}\mathcal{H}(\vecb{w})}
\,\Theta(\Delta)\,\Delta^{f-1} 
\end{equation}
for $r=0$, and
\begin{align}
&\bar{\rho}_{\vecb{w}}(\Delta)=
\left(\tfrac{1}{2\pi\hbar}\right)^{f}
\!\!\!
\tfrac{\sigma_{r}\sigma_{s}}{rs\det^{1/2}\mathcal{H}(\vecb{w})}
\frac{\partial}{\partial\Delta}
\biggl[\Delta^{\frac{s}{2}}\times
\nonumber\\
&{}_{2}F_{1}\bigl(\tfrac{r}{2},-\tfrac{s}{2},1+\tfrac{r}{2};-\tfrac{R_-^{2}}{\Delta}\bigr)
_{R_-=\Theta(-\Delta)|\Delta|^{1/2}}
^{R_-=e^{1/2}}
\biggr]
\label{eq:rhoGen}
\end{align}
for $r\neq0$.
Here, $\sigma_{d}=2\pi^{d/2}/\Gamma(d/2)$ with $d=r,s$ is the surface area of a $d$-dimensional ball with unit radius, ${}_{2}F_{1}(a,b,c;z)$ is the hypergeometric function of variable $z$ depending on parameters $\{a,b,c\}$, and $\Theta(x)$ is the step function ($\Theta\!=\!0$ for $x\!<\!0$ and $\Theta\!=\!1$ for $x\!\geq\!0$).
The Hessian determinant appears due to the $\vecb{x}\mapsto\vecb{y}$ transformation.
In the derivation of the above formulas, an integration in the subspace of positive Hessian eigenvalues has natural limits of $R_+=0$ and $\Theta(\Delta)|\Delta|^{1/2}$, while integration in the subspace of negative eigenvalues is bounded by the values of $R_-$ given in Eq.\,\eqref{eq:rhoGen}.
The parameter $e$ defines an artificial size of the stationary-point neighborhood for the evaluation of the volume function associated with the irregular part $\bar{\rho}_{\vecb{w}}$; its value is arbitrary in the sense that its change does not affect the qualitative results derived from the above expressions.

Let us outline these results.
For $r$ {\em even}, including $r=0$, an explicit form of the hypergeometric function yields the following formula:
\begin{equation}
\label{eq:reven}
\bar{\rho}_{\vecb{w}}(\Delta)\!=\!f_1(\Delta)\!+\!
\tfrac{\mathcal{C}_{1}\hbar^{-f}}{\det^{1/2}\mathcal{H}(\vecb{w})}\,(-)^{\frac{r}{2}}
\,\Theta(\Delta)\,\Delta^{f-1},
\end{equation}
where $f_1$ is a non-singular function (it collects all non-singular terms of the resulting expression and may be added to the regular part of the level density) and $\mathcal{C}_{1}$ is a positive constant.
We observe that the smooth level density and its derivatives up to $(f$$-$$2)$\,th one are continuous at $\Delta=0$, but the presence of the step function in Eq.\,(\ref{eq:reven}) implies that the $(f$$-$$1)$\,th derivative has 
a \emph{discontinuity}:
\begin{equation}
\label{eq:revend}
\frac{d^{f-1}\bar{\rho}_{\vecb{w}}}{dE^{f-1}}
\propto (-)^{\frac{r}{2}}\,\Theta(\Delta)
\,.
\end{equation}
The step of this derivative is oriented in the upward or downward direction for $r/2$ even or odd, respectively.
For $r$ {\em odd}, we obtain the following formula:
\begin{align}
\bar{\rho}_{\vecb{w}}(\Delta)&
=f_2(\Delta)+
\nonumber\\
&
\tfrac{\mathcal{C}_{2}\hbar^{-f}}{\det^{1/2}\mathcal{H}(w)}(-)^{\frac{r+1}{2}}\Delta^{f-1}\ln\abs{\Delta}
\,,
\end{align}
where again $f_2$ is a non-singular function and $\mathcal{C}_{2}$ a positive constant.
As in the previous case, the level density is continuous up to the $(f$$-$$2)$\,th derivative and the non-analyticity shows up in the $(f$$-$$1)$\,th derivative, which now exhibits a \emph{logarithmic divergence}:
\begin{equation}
\frac{d^{f-1}\bar{\rho}_{\vecb{w}}}{dE^{f-1}}
\propto (-)^{\frac{r+1}{2}}\ln\abs{\Delta}
\,.
\end{equation}
The diverging peak points downwards or upwards for $(r+1)/2$ even or odd, respectively.

Expressing the index of a stationary point as $r=4k+m$, where $k$ is an arbitrary integer, one can summarize the above findings on the singularities in $d^{f-1}\bar{\rho}/dE^{f-1}$ in terms of the value of integer $m$: 
(i) $m=0$ means an upward jump, 
(ii) $m=1$ implies a logarithmic divergence pointing upwards,
(iii) $m=2$ causes an downward jump, and 
(iv) $m=3$ results in a logarithmic divergence pointing downwards.
Assuming that in physically plausible situations $r\leq f$ (at each phase-space point, there exists $f$ independent kinetic terms that {\em increase\/} the total energy), we conclude that all four types of singularity of the respective level-density derivative appear for $f\geq 3$.  

Both expressions \eqref{eq:rho0} and \eqref{eq:rhoGen} can be easily extended also to fractional values of the number $f$.
Such an extension is not just a mathematical game---it is useful in systems in which the level density is obtained via integration of the form \eqref{eq:rhobar} in a space of \emph{odd dimension} $n$.
For instance, in lattice systems the density of vibrational or single-particle states is given by an integral of a particular dispersion relation over the momentum space \cite{Kan04}.
An odd spatial dimension $n$ then formally corresponds to a half-integer number $f=n/2$, which of course no more represents the true number of degrees of freedom. 
Alternatively, the extension to half-integer $f$ numbers may be relevant also in systems whose Hamiltonian depends explicitly on time~\cite{Rei04}.

The discussion for fractional $f$ is again split to the cases with $r$ even and odd.
For $r$ even, Eq.~\eqref{eq:reven} remains valid, but Eq.~\eqref{eq:revend} must be modified as follows:
\begin{equation}
\frac{d^{\lceil f-1\rceil}\bar{\rho}_{\vecb{w}}}{dE^{\lceil f-1\rceil}}
\propto (-)^{\frac{r}{2}}\,\Theta(\Delta)\,|\Delta|^{-\frac{1}{2}}
\,,
\end{equation}
where $\lceil x\rceil$ denotes the ceiling function, which in the present case yields $\lceil f$$-$$1\rceil=f$$-$$1/2$.
For $r$ odd, the irregular part of the level density and its $\lceil f$$-$$1\rceil$\,th derivative read as
\begin{align}
\bar{\rho}_{\vecb{w}}(\Delta)
&
=f_3(\Delta)+
\nonumber\\
&\tfrac{\mathcal{C}_{3}\hbar^{-f}}{\det^{1/2}\mathcal{H}(\vecb{w})}(-)^{f-\frac{r}{2}}
\,\Theta(-\Delta)\,|\Delta|^{f}
\,,
\end{align}
\begin{equation}
\frac{d^{\lceil f-1\rceil}\bar{\rho}_{\vecb{w}}}{dE^{\lceil f-1\rceil}}
\propto (-)^{\frac{r-1}{2}}\Theta(-\Delta)\,|\Delta|^{-\frac{1}{2}}
\,,
\end{equation}
where, as usually, $f_3$ is a non-singular function and $\mathcal{C}_{3}$ a positive constant.

It should be pointed out that the so-called Van Hove singularities in the eigenmode spectra of crystals with two or three spatial dimensions follow precisely the above classification \cite{Van53}.
At present, singularities of this type are studied in lattice systems like graphene in terms of ESQPTs \cite{Die13,Mac15}.

\subsection{Singularities caused by degenerate stationary points: Example}
\label{sec:dg}

A stationary point $\vecb{w}$ is degenerate if the Hessian matrix $\mathcal{H}(\vecb{w})$ has at least one eigenvalue equal to zero, \emph{i.e.}, if there exists a direction through $\vecb{w}$ in which both first and second derivatives of the Hamiltonian vanish.
In other words, the stationary point is \lq\lq flatter\rq\rq\ than quadratic in that direction. 
In such cases, one cannot in general expand the Hamiltonian near $\vecb{w}$ into a separable function as in Eq.\,(\ref{eq:HMorse}) and the Morse theory cannot be applied.

Discussion of the role of degenerate stationary points is often marginalized.
It is argued that even an infinitesimally small perturbation of the Hamiltonian transforms a degenerate stationary point into a non-degenerate one.
Indeed, degenerate stationary points generate structurally unstable functions of the catastrophe theory \cite{Dem00}, which under various perturbations change into different types of stable Morse functions (allowing a separable local expansion into quadratic functions in each point).
The structurally unstable functions represent only isolated points in the space the space of all smooth functions, while the Morse functions form an open dense subset in this space \cite{Dem00}.
Nevertheless, no matter how improbable is the occurrence of a degenerate stationary point in a real system, its potential impact on the level density can be much stronger than the effects of non-degenerate stationary points.

The catastrophe theory provides a classification of degenerate stationary points for problems of very low dimensionality.
However, for a general number of degrees of freedom $f$, an analysis of ESQPTs caused by degenerate stationary points has to be performed case by case.
Consider for example a separable local minimum of a general order.
The Hamiltonian close to the minimum can be cast in the form:
\begin{equation}
\label{eq:HDeg}
H(\vecb{y})=E_{\vecb{w}}+\sum_{i=1}^{2f}y_i^{m_i}
\,,
\end{equation}
where $\vecb{y}$ is a suitable coordinate system in the phase space centered in the minimum at $\vecb{w}$, and $m_i$ are positive powers associated with the components of $\vecb{y}$. 
Note that the function in Eq.\,\eqref{eq:HDeg} is analytic only when all $m_i$'s are even integers.
This locally separable type of Hamiltonian allows one to perform the integration in Eq.\,\eqref{eq:rhobar} explicitly. 
The resulting contribution to the irregular part of the level density is:
\begin{equation}
\label{eq:rhoDeg}
\bar{\rho}_{\vecb{w}}(\Delta)=\mathcal{C}_{4}\Theta(\Delta)\Delta^{g-1}
\,,\quad
g=\sum_{i=1}^{2f}\frac{1}{m_i}
\,.
\end{equation}
$\mathcal{C}_{4}$ is a positive constant given by:
\begin{equation}
\label{eq:C4}
\mathcal{C}_{4}=\left(\tfrac{2}{\pi\hbar}\right)^{f}\!\!
\tfrac{g}{\Gamma\left(1+g\right)}\tfrac{1}{{\rm det}{\cal J}}
\prod_{i=1}^{2f}\Gamma\left(1+\tfrac{1}{m_i}\right)
\,,
\end{equation}
where $\Gamma$ is the Euler gamma function and ${\rm det}{\cal J}$ a Jacobian of the $\vecb{x}\mapsto\vecb{y}$ transformation.

Analyzing properties of the expression~\eqref{eq:rhoDeg}, we pick up the following observations:
(i) The level density is discontinuous in the $\lceil g$$-$$1\rceil$\,th derivative at $E=E_{\vecb{w}}$, while all lower derivatives are continuous.
(ii) If $m_i=2$ $\forall i$ (that is, for a non-degenerate local minimum), we have $g=f$ and the level density behaves in accord with the previous analysis in terms of the Morse theory, see Eq.\,\eqref{eq:rho0}. 
(iii)	For a discontinuity occurring first in the $k$\,th derivative of the level density, the condition $g\in[k+1,k+2)$ must be fulfilled.
For instance, it happens if $2f/(k+2)< m_i \leq 2f/(k+1)$ for all $i$.
This means that for an increasing number of degrees of freedom, the stationary point needs to become increasingly flat if the associated non-analyticity should be observed in the level-density derivative of a given order.
(iv) For Hamiltonians of the standard form \eqref{eq:Hstand}, with a quadratic kinetic energy and an arbitrary potential, the powers in the momentum terms are fixed: $m_i=2$ for $i=1,\dots f$. 
In this case, even for \lq\lq infinitely flat\rq\rq\ potential with $m_i\rightarrow\infty$ for $i=(f\!+\!1),\dots 2f$, the discontinuity appears not sooner than in the $\lceil f/2-1\rceil$\,th derivative of the level density.

\section{Level dynamics}
\label{sec:Flora}

The ESQPTs are often investigated in systems whose Hamiltonians depend on some parameters.
Such a parameter, here denoted as $\lambda$, may represent an external control of the system (like a field strength) or an internal coupling constant (interaction strength between various parts of the system).
The dependence of the quantum spectrum on $\lambda$ (\lq\lq level dynamics\rq\rq) constitutes a very important aspect of an ESQPT that needs to be analyzed.

Besides the smoothed level density $\bar{\rho}(\lambda,E)$, which now depends on two variables, we introduce also a quantity $\bar{\phi}(\lambda,E)$ called the \emph{flow rate}~\cite{Str14}. 
It represents a smoothed slope of the spectrum at any point of the $\lambda\times E$ plane and can be derived from the continuity equation:
\begin{equation}
\label{eq:continuity}
\frac{\partial}{\partial\lambda}\bar{\rho}(\lambda,E)
+\frac{\partial}{\partial E}\left[\bar{\rho}(\lambda,E)\bar{\phi}(\lambda,E)\right]=0
\,.
\end{equation}
Here, $\bar{\phi}$ plays the role of a velocity field for the \lq\lq fluid\rq\rq\ of quantum states.
The continuity equation guarantees that the integral $\int\bar{\rho}(\lambda,E)\,dE$ over the whole spectrum is conserved as $\lambda$ changes.
If the smoothed level density is evaluated by convolution of the discrete energy spectrum with a smoothing function $F$, 
\begin{equation}
\bar{\rho}(\lambda,E)
=\sum_l F(E-E_l(\lambda))
\,,
\end{equation}
the flow rate can be extracted directly from the slopes of individual levels, locally averaged by the same smoothing function \cite{Str14}:
\begin{equation}
\label{eq:flora}
\bar{\phi}(\lambda,E)=\frac{1}{\bar{\rho}(\lambda,E)}
\sum_l F(E-E_l(\lambda))\frac{dE_l(\lambda)}{d\lambda}
\,.
\end{equation}
The flow rate can also be determined by solving Eq.\,\eqref{eq:continuity}, which leads to:
\begin{align}
\bar{\phi}(\lambda,E)=\frac{1}{\bar{\rho}(\lambda,E)}
&\biggl[\bar{\rho}(\lambda,E_{0})\bar{\phi}(\lambda,E_{0})
-
\nonumber\\
&\int_{E_{0}}^{E}dE'\frac{\partial\bar{\rho}}{\partial\lambda}(\lambda,E')\biggr]
\,,
\label{eq:phi}
\end{align}
where $E_{0}$ is an arbitrary energy point at which $\bar{\rho}$ is continuous.
Both expressions \eqref{eq:flora} and \eqref{eq:phi} are equivalent.

A natural question is how the existence of a stationary point of the Hamiltonian affects the analyticity of the flow rate $\bar{\phi}$.
If the Hamiltonian depends smoothly on $\lambda$, its stationary points (if any) form smooth \emph{critical borderlines} of ESQPTs in the plane $\lambda\times E$.
Let us assume that such a borderline is determined as a locus of points where the condition $h(\lambda,E)=0$ is satisfied for a certain smooth function $h$ (the \lq\lq quantum phases\rq\rq\ on both sides of the critical borderline yield values of $h$ with opposite signs).
The discontinuity or another type of non-analyticity of both $\bar{\rho}$ and $\bar{\phi}$ should be independent of the direction of crossing this borderline.
Differentiating Eq.\,\eqref{eq:continuity} with respect to energy $k$ times, we arrive at the following relation that must be satisfied at any point $(\lambda,E)$:
\begin{align}
\label{eq:kder}
&\underbrace{\frac{\partial}{\partial\lambda}\frac{\partial^k\bar{\rho}}{\partial E^k}
+\bar{\phi}\frac{\partial^{k+1}\bar{\rho}}{\partial E^{k+1}}}
_{\left[\vecb{\nabla}_{2}\,(\partial^k\bar{\rho}/\partial E^k)\right]\,\cdot\,\vecb{v}}
+\bar{\rho}\frac{\partial^{k+1}\bar{\phi}}{\partial E^{k+1}}
\\
&+\!\!\sum_{l=1}^{k}\!\binom{k}{l}\!\!
\left(\frac{\partial^{l}\bar{\phi}}{\partial E^{l}}\frac{\partial^{k-l+1}\bar{\rho}}{\partial E^{k-l+1}}
\!+\!\frac{\partial^{l}\bar{\rho}}{\partial E^{l}}\frac{\partial^{k-l+1}\bar{\phi}}{\partial E^{k-l+1}}\right)
\!=\!0.
\nonumber
\end{align} 
$\vecb{\nabla}_{2}\equiv(\frac{\partial}{\partial\lambda},\frac{\partial}{\partial E})$ stands for the gradient in the $\lambda\times E$ plane and $\vecb{v}\equiv(1,\bar{\phi})$ for a vector in this plane pointing in the direction of the averaged flow of the spectrum.
With this notation it becomes clear that the sum of the first two terms in Eq.\,\eqref{eq:kder} is proportional to the derivative of $(\partial^k\bar{\rho}/\partial E^k)$ in the direction of $\vecb{v}$.

>From the expression \eqref{eq:phi} it follows, for $\bar{\rho}\neq 0$ (which we assume below), that the continuity of $\bar{\rho}$ at any point $(\lambda,E)$ implies the continuity of $\bar{\phi}$ at that point, and \emph{vice versa}. 
By using Eq.\,\eqref{eq:kder} successively with increasing order of derivative we can see that 
continuity of the derivatives of $\bar{\rho}$ up to order $k$ is equivalent with continuity of 
the $\bar{\phi}$ derivatives up to the same order $k$.
Now, if the $(k$$+$$1)$\,th derivatives of $\bar{\rho}$ is discontinuous or divergent at the critical borderline $h(\lambda,E)=0$, the sum of the first two terms in Eq.\,\eqref{eq:kder} will behave in the same way if the flow direction $\vecb{v}$ is not parallel with this borderline.
Therefore, under this condition, the same type of non-analyticity must affect also the $(k$$+$$1)$\,th energy derivative of $\bar{\phi}$, otherwise the relation \eqref{eq:kder} would fail.
Of course, if the flow direction $\vecb{v}$ is parallel with the critical borderline, the flow rate does not have to develop any type of non-analyticity (as can be illustrated by straightforward examples), but this is not a generic case.

In summary, we come to the conclusion that in generic situations the flow rate $\bar{\phi}$ of the quantum spectrum at the ESQPT critical borderline exhibits the same type of non-analyticity as the level density $\bar{\rho}$.
Hence the analysis from Sec.\,\ref{sec:Den} applies also to the flow rate.
It should be pointed out that the signs of singularities (the jump or logarithmic divergence) in $\bar{\rho}$ and $\bar{\phi}$ can differ due to the fact that the flow rate, in contrast with level density, allows for both positive and negative values.

\section{A toy model with $\vecb{f}$\,=\,3}
\label{sec:Model}

The ESQPTs in systems with $f=1$ and $2$ degrees of freedom, described by Hamiltonians of the type \eqref{eq:Hstand}, have been already analyzed in detail~\cite{Cej08,Str14}.
The results can be summarized as follows:
(i) The $f=1$ case offers two types of non-degenerate stationary points in the potential: local minimum, $r=0$, which manifests itself as an upward jump in $\bar{\rho}$, and local maximum, $r=1$, which gives rise to a logarithmic divergence in $\bar{\rho}$. 
(ii) The $f=2$ case has three different types of non-degenerate stationary points in the potential.
They all show up as singularities in the first derivative $\partial\bar{\rho}/\partial E$: a local minimum, $r=0$, causes an upward jump, a saddle point, $r=1$, produces a logarithmically divergent peak, and a local maximum, $r=2$, leads to a downward jump.
These results are in perfect agreement with the above-derived general conclusions.

Here we give an example of multiple ESQPTs in a simple system with $f=3$ (the lowest dimension in which all four singularity types from Sec.\,\ref{sec:ndg} can be observed). 
The Hamiltonian is composed of three independent components, each of them written as:
\begin{equation}
\label{eq:cusp}
H_{{\rm 1D}}(p,q;A)=\frac{p^{2}}{2}+Aq-2q^2+q^4\,,
\end{equation}
where $q$ is the coordinate, $p$ the corresponding momentum, and $A$ a tunable parameter.
The potential has two locally quadratic minima separated by a local quadratic maximum for $\abs{A}<\sqrt{64/27}\doteq 1.54$ (both minima having the same energy for $A=0$), and only one quadratic minimum outside this interval.
Note that the potential from Eq.\,\eqref{eq:cusp} is known from the context of the catastrophe theory as the \lq\lq cusp\rq\rq\ potential (the quartic term is a germ of the cusp catastrophe) \cite{Dem00}.

\begin{table}[h!]
\begin{tabular}{|c|c|c|c|c|}
\hline
\# & $E$      & $q_1,q_2,q_3$        & type         & $r$ \\
\hline
 1 & $-4.551$ & $-1.03,-1.06,-1.08$  & min          & 0 \\
 2 & $-4.051$ & $+0.97,-1.06,-1.08$  & min          & 0 \\
 3 & $-3.553$ & $-1.03,+0.93,-1.08$  & min          & 0 \\
 4 & $-3.289$ & $+0.06,-1.06,-1.08$  & sad           & 1 \\
 5 & $-3.058$ & $-1.03,-1.06,+0.89$  & min          & 0 \\
 6 & $-3.053$ & $+0.97,+0.93,-1.08$  & min          & 0 \\
 7 & $-3.005$ & $-1.03,+0.13,-1.08$  & sad           & 1 \\
 8 & $-2.697$ & $-1.03,-1.06,+0.20$  & sad           & 1 \\
 9 & $-2.558$ & $+0.97,-1.06,+0.89$  & min          & 0 \\ 
10 & $-2.505$ & $+0.97,+0.13,-1.08$  & sad           & 1 \\
11 & $-2.291$ & $+0.06,+0.93,-1.08$  & sad           & 1 \\
12 & $-2.197$ & $+0.97,-1.06,+0.20$  & sad           & 1 \\
13 & $-2.060$ & $-1.03,+0.93,+0.89$  & min          & 0 \\
14 & $-1.796$ & $+0.06,-1.06,+0.89$  & sad           & 1 \\ 
15 & $-1.743$ & $+0.06,+0.13,-1.08$  & sad           & 2 \\
16 & $-1.699$ & $-1.03,+0.93,+0.20$  & sad           & 1 \\
17 & $-1.560$ & $+0.97,+0.93,+0.89$  & min          & 0 \\
18 & $-1.512$ & $-1.03,+0.13,+0.89$  & sad           & 1 \\
19 & $-1.435$ & $+0.06,-1.06,+0.20$  & sad           & 2 \\
20 & $-1.199$ & $+0.97,+0.93,+0.20$  & sad           & 1 \\
21 & $-1.151$ & $-1.03,+0.13,+0.20$  & sad           & 2 \\
22 & $-1.012$ & $+0.97,+0.13,+0.89$  & sad           & 1 \\
23 & $-0.798$ & $+0.06,+0.93,+0.89$  & sad           & 1 \\
24 & $-0.651$ & $+0.97,+0.13,+0.20$  & sad           & 2 \\
25 & $-0.437$ & $+0.06,+0.93,+0.20$  & sad           & 2 \\
26 & $-0.250$ & $+0.06,+0.13,+0.89$  & sad           & 2 \\
27 & $+0.111$ & $+0.06,+0.13,+0.20$  & max          & 3 \\
\hline        
\end{tabular}
\caption{Energies (ordered from the lowest to the highest), coordinates, and types (including Morse index $r$) of the 27 stationary points of the triple cusp Hamiltonian \eqref{eq:CUSP3D}.}
\label{tab:sp}
\end{table}

The Hamiltonian of the $f=3$ system is then just a sum of three 1D cusp terms:
\begin{equation}
\label{eq:CUSP3D}
H_{{\rm 3D}}(\vecb{p},\vecb{q};\vecb{A})=\sum_{i=1}^{3}H_{{\rm 1D}}(p_i,q_i;A_i)
\end{equation}
where the \lq\lq vector\rq\rq\ of parameters is chosen as $\vecb{A}\equiv(A_1,A_2,A_3)=(1/4,1/2,3/4)$ so that each of the 1D potentials has 3 stationary points.
Hence the full  Hamiltonian~\eqref{eq:CUSP3D} has in total $3^3=27$ stationary points (including the global minimum); their list is given in Table~\ref{tab:sp}.
All these stationary points are non-degenerate.
Note that the abundance of stationary points in the present $f=3$ model illustrates the fact that the number of stationary points is generally expected to grow exponentially with the number of degrees of freedom $f$. 

\begin{figure*}[t]
\begin{flushright}
\epsfig{file=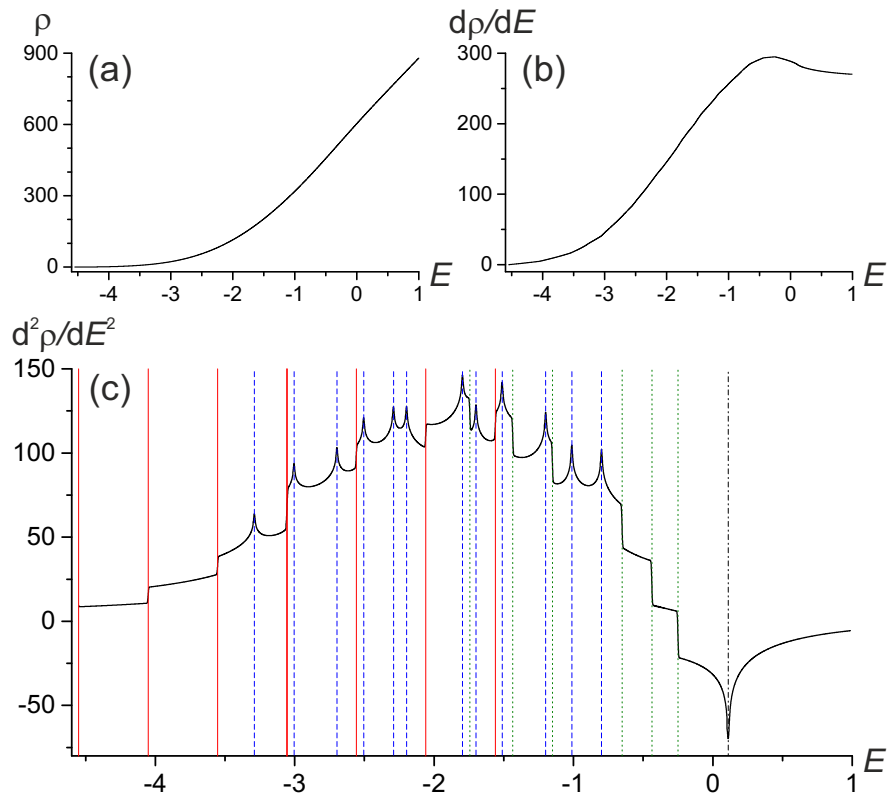,width=0.7\linewidth}
\caption{
(Color online)
Level density, normalized by $(2\pi\hbar)^{3}$, and its first and second derivatives for the triple cusp Hamiltonian~\eqref{eq:CUSP3D} with $\hbar=5\cdot 10^{-4}$. 
The ESQPTs of different types, corresponding to stationary points from Tab.\,\ref{tab:sp}, are marked with vertical lines: minimum $r=0$ (red solid line), saddle point $r=1$ (blue dashed line), saddle point $r=2$ (green dotted line), and maximum $r=3$ (black dot-dashed line).
The figure was obtained by drawing approximately $10^{11}$ energy levels from the numerical diagonalization into a normalized histogram with $10^4$ equidistant bins.}
\label{fig:sp}
\end{flushright}
\end{figure*}

The level density and its first and second derivatives, obtained via smoothing of a quantum spectrum from numerical diagonalization of the Hamiltonian \eqref{eq:CUSP3D}, are shown in Figure~\ref{fig:sp}.
The level density is plotted in panel (a) and the first and second derivatives in panels (b) and (c), respectively.
Both level density and its first derivative are apparently continuous functions of energy.
Indeed, one would hardly anticipate any kind of non-analyticity if looking only at panel (a).
However, the second derivative in panel (c) clearly discloses all 26 stationary points above the global minimum, as listed in Tab.\,\ref{tab:sp}.
We see that all local minima ($r=0$) induce upward jumps of $\partial^2\bar{\rho}/\partial E^2$, while the saddle points with $r=1$ give rise to logarithmic divergences pointing upwards and the saddle points with $r=2$ lead to downward jumps.
Finally, the only local maximum ($r=3$) yields a logarithmic divergence pointing downwards.
These observations are in perfect agreement with the theoretical predictions developed in Sec.\,\ref{sec:ndg}.

\begin{figure*}[t]
\epsfig{file=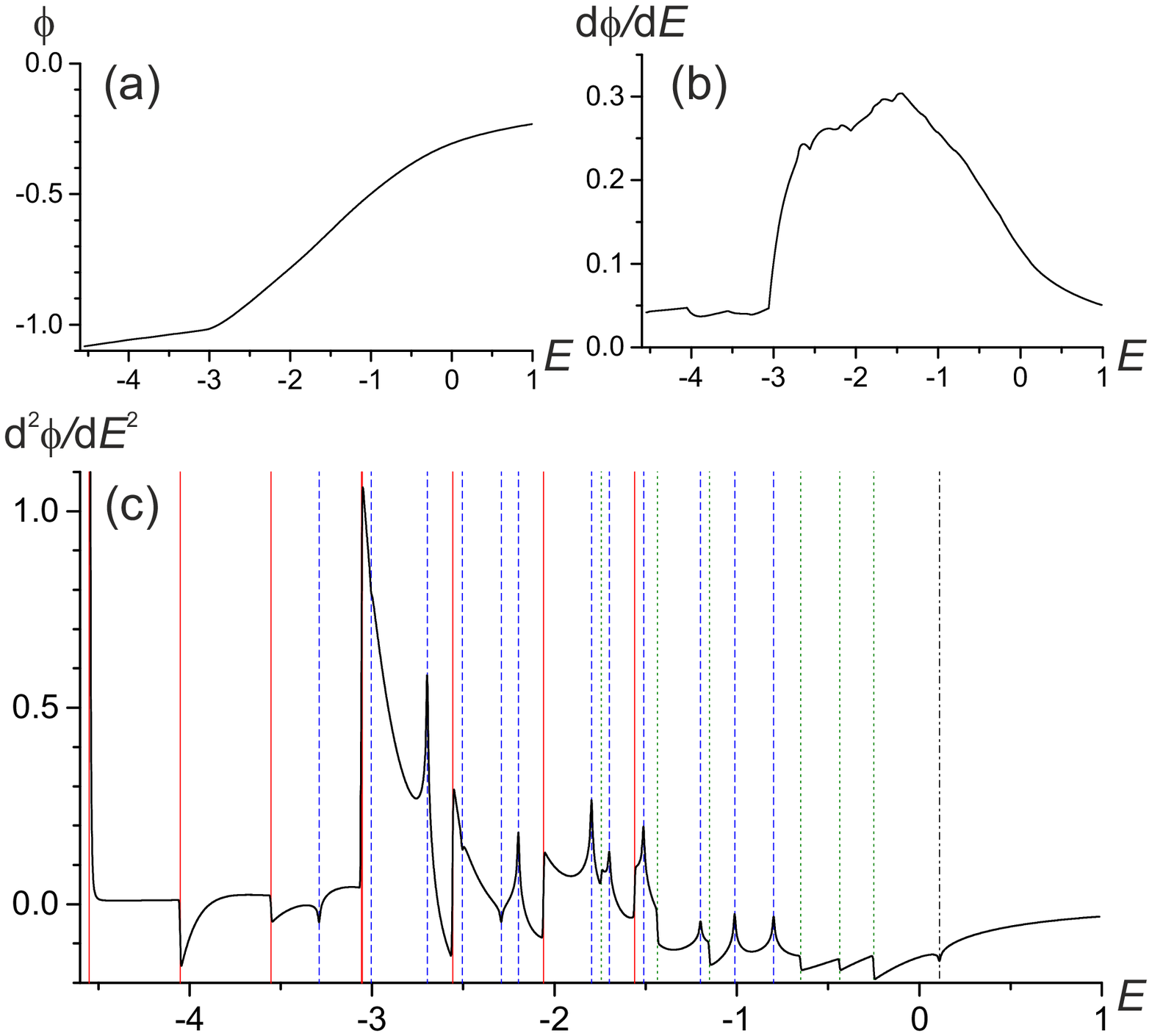,width=0.7\linewidth}
\caption{
(Color online)
The same as in Fig.\,\ref{fig:sp} but for the flow rate \eqref{eq:phi} induced by an infinitesimal variation of parameter $A_3$ of Hamiltonian \eqref{eq:CUSP3D}.
The slopes of individual levels were obtained by a numerical differentiation.
The observed singularities are related to those in the level density, in agreement with the analysis in Sec.\,\ref{sec:Flora}. 
}
\label{fig:fl}
\end{figure*}

Figure \ref{fig:fl} shows the flow rate of the spectrum for variable parameter $\lambda\equiv A_3$.
It is evaluated from Eq.\,\eqref{eq:phi} by a numerical differentiation of individual level energies at the same parameter vector $\vecb{A}$ as used above.
At the known stationary point energies the flow rate shows the same types of non-analyticities (explicit in the second derivative of $\bar{\phi}$ with respect to energy) as the level density.
Note that the signs of individual jumps/divergences differ from those observed in Fig.\,\ref{fig:sp}.
All these results verify conclusions of Sec.\,\ref{sec:Flora}.

\section{Summary}
\label{sec:Conclusion}

We have studied non-analyticities in quantal spectra induced by classical stationary points of a general Hamiltonian.
The non-analyticities appear in the semiclassical density of quantum energy levels, and---if the Hamiltonian depends on a tunable control parameter---also in the so-called flow rate, {\it i.e.}, the smoothed slope of the spectrum as a function of the parameter.
Although real non-analyticities appear only in the system's infinite-size limit, which in the models with finite numbers of degrees of freedom $f$ coincides with the classical limit, $\hbar\to 0$, rather sharp precursors of this behavior can observed already in finite size cases.

We have provided a classification of such non-analyticities for non-degenerate (fully quadratic) stationary points.
The classification is given by a pair of numbers $(f,r)$, where $f$ is the number of degrees of freedom (or a half of this number if the semiclassical level density is obtained by integration only in the momentum space---like in lattice systems) and $r$ is the index of the stationary point (the number of negative eigenvalues of the Hessian matrix associated with the Hamiltonian).
We found that if $f$ is an integer, the stationary point of the type $(f,r)$ affects the $(f$$-$$1)$\,th derivative of the smooth level density, which exhibits either a jump (for $r$ even) or a logarithmic divergence (for $r$ odd).
The jump as well as the divergence can be oriented upwards or downwards, depending on the value $m=r({\rm mod}\,4)$.
These conclusions were demonstrated in the smoothed level density of a toy model with $f=3$.
If $f$ is a half-integer (the semiclassical level density results from integration in a space of odd dimension), the $(f,r)$ stationary point shows up in the $\lceil f$$-$$1\rceil\equiv (f-1/2)$\,th derivative of the level density, which exhibits an inverse square root singularity on one side of the critical energy.

In contrast, a general classification of spectral singularities associated with degenerate (non-quadratic) stationary points is not available as there is no analog of the Morse lemma for such points.
Therefore, the effects of degenerate stationary points must be studied case by case. 
We have considered an example of a separable local minimum, with the potential along individual axis directions varying locally according to power law dependences with arbitrary degrees.
We have seen that the non-analyticity appears in lower derivatives of the semiclassical level density if the degrees of the power law dependences increase, {\it i.e.}, if the minimum gets ``flatter''.

It has been shown that under generic conditions, the smoothed flow rate of the spectrum at the ESQPT critical borderline exhibits the same type of non-analyticity as the smoothed level density.
An exception from this rule is possible if the local flow of the spectrum is parallel with the critical borderline.
The correspondence between the flow-rate and level-density singularities was verified in the $f=3$ toy model. 

The general conclusions of our analysis are now ready for use in specific quantum systems that exhibit ESQPT effects in arbitrary numbers of degrees of freedom.
Finite algebraic models describing collective behavior of interacting $N$-particle systems are particularly  suitable for such applications.
These models in their infinite-size limit $N\sim 1/\hbar\to\infty$ typically yield Hamiltonians that couple coordinates and momenta in a rather non-trivial way, see {\it e.g.} Ref.\,\cite{Mac14}.
Therefore, the outcomes of the present work represent an adequate basis for the ESQPT analysis in such systems.

\section*{Acknowledgments}

The authors acknowledge enlightening discussions with M. Macek, F. Iachello, and M. Kastner.
This work was performed under the project no. P203-13-07117S of the Czech Science Foundation.

\end{document}